\documentclass{mn2e}
\usepackage{epsf}
\usepackage{graphicx}

\def\etal{{\it et~al. }}

\def\mdot{{M_{\odot}}}
\def\mdo{{\dot m}}

\title[Radio observations of Circinus X-1 over a complete orbit]
{Radio observations of Circinus X-1 over a complete orbit at an historically faint epoch}

\author[D.E. Calvelo \etal]
{D.E. Calvelo,$^{1}$ R.P. Fender,$^{1}$ A.K. Tzioumis,$^{2}$ N. Kawai, $^{3}$ J.W. Broderick, $^{1}$ 
\newauthor M.E. Bell $^{1}$ \\
$^{1}$ School of Physics and Astronomy, University of Southampton,
Highfield, Southampton, SO17 1BJ, UK\\
$^{2}$ Australia Telescope National Facility, CSIRO, PO Box 76, Epping, New south Wales 1710, Australia\\
$^{3}$ Department of Physics, Tokyo Institute of Technology, 2-12-1 Ohokayama, Meguro, Tokyo 152-8551, Japan\\}

\date{\today}

\volume{000}

\pagerange{\pageref{firstpage}--\pageref{lastpage}} \pubyear{2010}

\begin{document}

\label{firstpage}

\maketitle
\begin{abstract}

We present results from the first radio observations of a complete
orbit ($\sim$ 17 days) of the neutron star X-ray binary Circinus X-1
using the Australia Telescope Compact Array Broadband Backend, taken
while the system was in an historically faint state. We have captured
the rapid rise and decline of a periastron passage flare, with flux
densities for 9 days prior to the event stable at $\sim$ 1 mJy at 5.5
GHz and $\sim$ 0.5 mJy at 9 GHz. The highest flux densities of 43.0 $\pm$ 0.5 mJy at 5.5
GHz and 29.9 $\pm$ 0.6 mJy at 9 GHz were measured during the flare's decline (MJD 55206.69) which continues towards pre-flare flux densities over the following 6
days. Imaging of pre-flare data reveals steady structure including two
stable components within 15 arc-seconds of the core which we believe
may be persistent emission regions within the system's outflows, one
of which is likely associated with the system's counter-jet. Unlike
past observations carried out in the system's brighter epochs, we observe no significant structural variations within $\approx$ 3
arc-seconds of the core's position. Model subtraction and difference
mapping provide evidence for variations slightly further from the
core: up to 5'' away. If related to the observed core flare, then
these variations suggest very high outflow velocities with $\Gamma$
$>$ 35, though this can be reduced significantly if we invoke phase delays of at least one orbital
period. Interestingly, the strongest structural variations appear to
the north west of the core, opposite to the strongest arcsec-scale
emission historically. We discuss the implications of this behaviour,
including the possibility of precession or a kinked approaching jet.

\end{abstract}


\section{Introduction}

\begin{table*}
\caption{Circinus X-1 observation log. The table lists the dates of observations, the Modified Julian Day (MJD) of the beginning of observations (on source), the total on-source time of each observing run, average daily radio flux densities at each frequency (5.5 GHz and 9 GHz, measured via point source fits) with the range of daily light curve values included in square brackets, along with image noise levels. The average flux densities become less useful after flare onset at the end of observations on 2010 Jan 09, due to the level of variation observed over single observation sessions (reflected in the uncertainties = daily light curve $\sigma$). It should be noted that that the values listed for 2010 Jan 09 are measured from data that does not include the flare section towards the end of observations.}
\begin{center}
\begin{tabular}{|l|c|c|c|c|c|c|c|}
Date & MJD & MJD & Total & F$_{5.5}$ [range] & Noise $_{5.5}$ & F$_{9}$ [range] & Noise $_{9}$ \\
(UT) & start & end & time (h) & (mJy beam$^{-1}$) & ($\mu$Jy beam$^{-1}$) & (mJy beam$^{-1}$) & ($\mu$Jy beam$^{-1}$) \\
\hline
2009 Dec 30 & 55195.745 & 55196.055 & 6.50 & 0.89 $\pm$ 0.08 [0.72:1.06] & 9.7 & 0.58 $\pm$ 0.11 [0.39:0.77] & 13.7 \\
2009 Dec 31 & 55196.707 & 55197.056 & 6.91 & 1.16 $\pm$ 0.11 [0.88:1.42] & 12.9 & 0.75 $\pm$ 0.10 [0.66:1.01] & 24.0 \\
2010 Jan 01 & 55197.756 & 55198.057 & 6.29 & 0.67 $\pm$ 0.15 [0.47:0.94] & 14.5 & 0.41 $\pm$ 0.14 [0.32:0.72] & 30.3 \\
2010 Jan 02 & 55198.704 & 55199.076 & 5.82 & 0.99 $\pm$ 0.07 [0.87:1.10] & 10.8 & 0.60 $\pm$ 0.08 [0.51:0.77] & 15.5 \\
2010 Jan 04 & 55200.706 & 55201.181 & 9.91 & 0.89 $\pm$ 0.08 [0.73:1.10] & 8.6 & 0.55 $\pm$ 0.09 [0.39:0.73] & 11.1 \\
2010 Jan 05 & 55201.703 & 55202.162 & 9.43 & 0.77 $\pm$ 0.17 [0.37:0.91] & 9.0 & 0.41 $\pm$ 0.18 [0.18:0.72] & 11.8 \\
2010 Jan 06 & 55202.703 & 55203.160 & 7.15 & 0.49 $\pm$ 0.12 [0.29:0.78] & 11.1 & 0.25 $\pm$ 0.06 [0.18:0.40] & 13.3 \\
2010 Jan 07 & 55203.682 & 55204.148 & 9.51 & 0.58 $\pm$ 0.11 [0.41:0.79] & 9.1 & 0.33 $\pm$ 0.10 [0.22:0.58] & 12.1 \\
2010 Jan 08 & 55204.687 & 55205.163 & 9.43 & 0.59 $\pm$ 0.10 [0.44:0.75] & 9.1 & 0.30 $\pm$ 0.09 [0.17:0.50] & 12.1 \\
2010 Jan 09 & 55205.681 & 55206.157 & 9.41 & 0.93 $\pm$ 0.42 [0.42:6.67] & 9.8 & 0.93 $\pm$ 0.65 [0.23:9.83] & 12.2 \\
2010 Jan 10 & 55206.683 & 55207.162 & 10.03 & 21.3 $\pm$ 8.81 [11.7:43.0] & 38.4 & 16.6 $\pm$ 6.10 [6.57:29.9] & 43.0 \\
2010 Jan 11 & 55207.684 & 55208.139 & 9.53 & 9.23 $\pm$ 1.33 [7.60:11.3] & 20.5 & 6.11 $\pm$ 1.26 [2.64:6.21] & 25.4 \\
2010 Jan 12 & 55208.661 & 55209.150 & 10.19 & 5.48 $\pm$ 0.80 [2.05:5.61] & 16.5 & 4.02 $\pm$ 0.69 [0.96:3.65] & 18.3 \\
2010 Jan 13 & 55209.685 & 55210.153 & 7.46 & 3.72 $\pm$ 0.51 [2.48:3.96] & 15.6 & 3.19 $\pm$ 0.21 [0.97:2.31] & 20.1 \\
2010 Jan 14 & 55210.672 & 55211.142 & 9.71 & 2.37 $\pm$ 0.21 [1.61:2.35] & 10.4 & 2.39 $\pm$ 0.13 [0.99:1.43] & 12.8 \\
2010 Jan 15 & 55211.665 & 55212.072 & 8.22 & 1.54 $\pm$ 0.11 [1.32:1.79] & 9.9 & 0.91 $\pm$ 0.12 [0.75:1.08] & 14.7 \\
\hline
\end{tabular}
\end{center}
\end{table*}

Circinus X-1 (Cir X-1) is a peculiar X-ray binary (XRB) system known
for its regular outbursts occurring every 16.6 days. These events are believed to be
caused by a highly eccentric orbit, wherein the periastron passage
results in an increased level of accretion onto the compact object
(Murdin \etal 1980; Nicolson, Glass \& Feast 1980). These outbursts
are visible at multiple wavelengths, including X-ray (Tennant, Fabian
\& Shafer 1986), IR (Glass 1978) and radio (Whelan \etal
1977). Discovered in 1971 (Margon \etal 1971), the system was
initially classified as a black hole candidate due to X-ray spectral
and timing similarities to Cygnus X-1, including millisecond
variability (Toor 1977). However, a reclassification was necessary
after the discovery of type I X-ray bursts in 1985 which indicated a
neutron star (NS) primary (Tennant, Fabian \& Shafer 1986), though
none had been seen again until May 2010, when the NS host was
confirmed by RXTE detection of bursts (Linares \etal 2010). Though the
system is often classified as a low mass XRB, the nature of the
companion star remains under debate, with the possibility of it being
a 3 - 5 $\mdot$ sub-giant (Johnston, Fender \& Wu 1999) or even a
super-giant up to 10 $\mdot$ (Jonker, Nelemans \& Bassa 2007).

Cir X-1 also challenges normal NSXRB classification in that it shows
X-ray behaviour reminiscent of both atoll (Oosterbroek \etal 1995) and
Z-source classes (Shirley, Bradt \& Levine 1997), as well as behaviour
that defies either classification in orbital phases prior to
periastron (Soleri \etal 2009a). This divide is likely related to
variations in accretion rate, $\mdo$, onto the neutron star during its
orbit, much like the different behaviours observed from XTE J1701-462
which have been linked to changes in $\mdo$ over time (Homan \etal
2007).

Low-frequency radio images reveal an extensive jet-powered nebula surrounding Cir X-1 (Stewart \etal 1993, Tudose \etal 2006). Early `runaway binary' theories, which suggested the system was associated with nearby supernova remnant G321.9-0.3 (Clark, Parkinson \& Caswell 1975) were supported by the existence of a tail-like structure extending from the southern end of Cir X-1's nebula. However, observations with the $Hubble$ $Space$ $Telescope$ revealed little or no proper motion, reducing the likelihood of such an origin (Mignani \etal 2002). 

Nearby (arcsec) ejecta, presumably previously expelled from Cir X-1, were found to brighten in the radio after flare events, suggesting re-energisation by highly relativistic invisible outflows with $\Gamma$ $>$ 15 (assuming association with the flare immediately preceding the change: Fender \etal 2004) and inclination angles close to the line of sight ($\theta$ $<$ 5$^{\circ}$). Cir X-1 was the second (after Sco X-1) neutron star system to show evidence of such invisible relativistic outflows. The implied velocities make the jets from Circinus X-1 some of the fastest in our Galaxy. These calculations assume a distance to the source of 6.5 kpc as put forward by Stewart \etal (1993), but that also serves as a compromise between the more recent estimates of Jonker \& Nelemans (2004: 7.8 - 10.5 kpc) and Iaria \etal (2005: 4.1 kpc). The jets themselves are resolved on arc-second scales in X-rays (Heinz \etal 2007, Soleri \etal 2009b) and arc-second to arc-minutes in radio (Stewart \etal 1993, Tudose \etal 2006), with the jets appearing curved on the larger scales. X-ray shocks, believed to be caused by jet impacts, have also been observed (Sell \etal 2010). There have been suggestions of jet precession occurring within the system, based on conflicting estimates for the inclination of the outflows (Iaria \etal 2008: $\theta$ $\approx$ 90$^{\circ}$).

Between 1997 and 2002 Cir X-1 had been growing steadily fainter in the X-rays, and though radio flare events were known to reach $>$ 1 Jy in the 1970s (Haynes \etal 1978) they too had been in decline, reaching at most 10s of mJy by the 2000s (Fender, Tzioumis \& Tudose 2005). Some renewed radio activity, similar to that observed in the 70s and 80s, occurred in 2006 (Nicolson 2007) but was followed by a return to minimal activity; a historically faint period during which our observations were taken. The observations and results discussed herein were taken during this most recent faint period. The system entered a relatively active phase in mid 2010 (a few months after our observations), with X-ray flares reaching hundreds of mCrab (Nakajima \etal 2010) and radio flares $>$ 0.1 Jy (Calvelo \etal 2010). Continued X-ray monitoring has shown the system intermittently returning to a low activity level for several months, much like that during which our observations were taken, and then undergoing another period of intense flaring. Thus, it seems these epochs of `low' activity are punctuated by short but distinct flaring episodes, which appear to be extreme examples of the periastron events.

\begin{figure*}
\centerline{\includegraphics[width=7.1in]{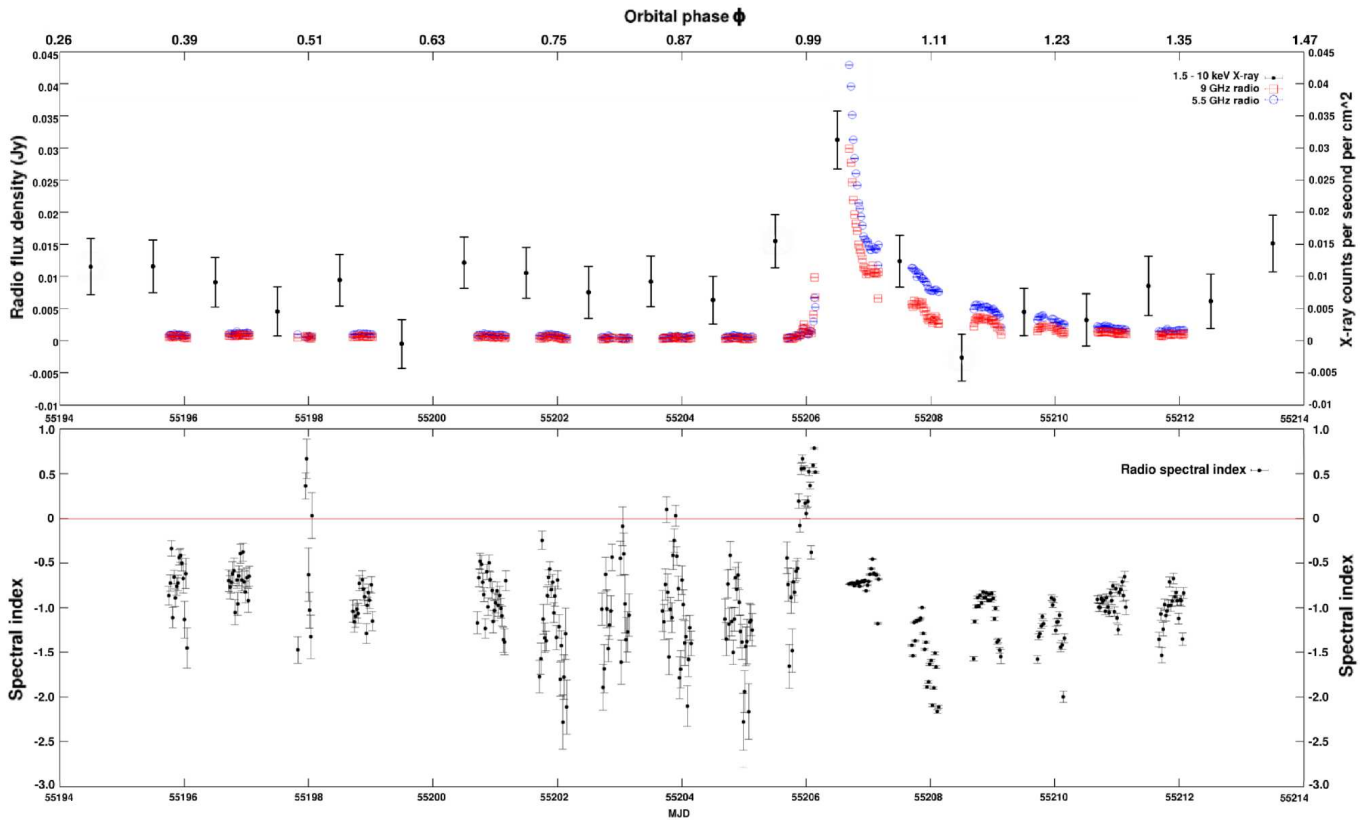}}
\caption{MAXI X-ray, ATCA-CABB radio (top panel) and radio spectral index (bottom panel) light curves for our observation period. Radio flux density errors (1$\sigma$) are included but are too small to distinguish clearly. The flare event is easily noticeable in all light curves, as well as the expected effects on spectral index during rise and decay.}
\end{figure*}

\section{Observations \& Data Reduction}

Our goal was to monitor Cir X-1 over a complete orbital period using the Australia Telescope Compact Array - Compact Array Broadband Backend (ATCA-CABB), with the hope of catching the flare in significant detail as well as its subsequent decline, and additionally, to see if the structure of the source varied in any way around an orbit (Moin et al. (2011) recently carried out a similar campaign on milli-arcsecond scales with eVLBI but were only able to detect the source shortly after periastron). Furthermore, with the Monitor of All-sky X-ray Image (MAXI: Matsuoka \etal 2009) monitoring the system in X-rays, we would be able to simultaneously compare a flare event in the two bands.

By continuing to observe after the flare we would be able to capture any variation in ejecta luminosities caused by unseen relativistic jets, or perhaps even the jets themselves. However, given the observation time available post-flare ($\sim$ 6 days), an assumed object distance of 6.5 kpc and a maximum image resolution of 1.5'', we would only be able to observe flows with minimum $\beta$ $\approx$ 9.5. Therefore, if we hope to observe the effects of flows related to our predicted flare we would require flow velocities similar to those implied by Fender et al. 2004. This does not prevent us from observing structural variations at any point during the run which might arise as a result of ejections from earlier events.

\subsection{Radio}

Observations of Cir X-1 were carried out on 2009 Dec 30, 31, 2010 Jan, 01, 02, 04, 05, 06, 07, 08, 09, 10, 11, 12, 13, 14 and 15 using the ATCA-CABB in 6A configuration (minimum baseline of 337m, maximum of 5939m) at both 5.5 GHz and 9 GHz (see Table 1). PKS J0825-5010 (PKS B0823-500) was used as the primary flux and bandpass calibrator when ever possible, with PKS J1939-6342 (PKS B1934-638) used in a few cases. Phase calibration was carried out using PMN J1515-5559 (PKS B1511-55) for the first half of the observations until it was noticed that on-screen levels indicated polarisation (5-10\%) of the source, at which point PMN J1524-5903 (PKS B1520-58) was made the phase calibrator, with some days including observations of both phase calibrators to allow us to eliminate the slight polarisation effect during analysis. Observation times varied between 8 and 12 hour runs each day (predicted RMS noise between 8 and 6 $\mu$Jy at 5.5 GHz and between 10 and 8 $\mu$Jy at 9 GHz), for a total of $\sim$ 135 hours on the source. Deconvolution was carried out using a combination of MFCLEAN (multi-frequency: Sault and Wieringa 1994) and original CLEAN (H\"{o}gbom 1974) subroutines. All data and image processing was carried out in MIRIAD (Sault, Teuben and Wright 1995).

\section{Analysis \& Results}

\subsection{Flare event}

Using the ephemeris detailed in Nicolson 2007 (ATEL \#985) we
predicted a flare would occur near 2010 Jan 11. Observations from 2009
Dec 30 through to 2010 Jan 09 indicated Circinus X-1 was relatively
stable at flux densities of $\sim$ 1 mJy at 5.5 GHz, and slightly
lower at 9 GHz (for a list of average daily flux densities see Table
1). A sudden rise in flux density was detected towards the end of a
day's observations at 2010 Jan 10 02:30 UT, continuing until scheduled
time ended at $\sim$ 03:50. Upon returning to the source $\sim$ 12.5
hours later, we found Cir X-1 in decline, indicating we had missed the
peak of the outburst.

To better analyse the flare we divided data-sets from each day into
smaller cuts ($\sim$ 30 minutes), produced images from these cuts, and
used the miriad command IMSTAT to measure the maximum flux density
within a 10$\times$10 arc-sec$^{2}$ box around the system's previously
established position (15:20:40.9 -57:10:00; based on image fitting from Tudose \etal 2008: tables A1 \& A2); i.e. the system's `core' from which we observe
the highest levels of radio emission. This allowed us to create a
detailed light curve (Figure 1) while maintaining an acceptable level
of error on each measurement.

The flare event (shown in more detail in Figure 2) appears to start at
MJD $\sim$ 55206.05, rising from pre-flare levels of 1 mJy to
S$_{5.5}$ $\sim$ 6.5 mJy and S$_{9}$ $\sim$ 10 mJy in just under 2
hours, before the day's observations finally ended. Though some
variations in flux densities are visible in pre-flare day light curves
(a scatter of $\sim$ 0.5 mJy visible in most) they do not follow any
obvious trend. However, the light curve from the day of the flare
shows a gradual rise in flux density prior to the event, with levels
at the start of the run $\sim$ 0.5 mJy. There is also a brief peak in
flux density at both frequencies at MJD 55205.97 (2 hours prior to the
flare - more easily visible in Figure 2), about one hour in
duration. Observations the following day began at MJD 55206.683; the
initial flux densities of S$_{5.5}$ = 43.0 $\pm$ 0.5 mJy and S$_{9}$ =
29.9 $\pm$ 0.6 mJy proceeded to decay rapidly (roughly power law
decay: log S$_{9GHz}$ $\approx$ -3.03(log T) - 2.12 at 9 GHz, log S$_{5.5GHz}$ $\approx$ -3.03(log T) - 1.96, where T is days since outburst start i.e. MJD - 55206.05) before appearing to level off at S$_{5.5}$ $\sim$ 15 mJy and
S$_{9}$ $\sim$ 11 mJy near MJD 55207. Subsequent days show continued
decay (once again a rough power law: log S$_{9GHz}$ $\approx$ -1.4(log T) - 1.9, log S$_{5.5GHz}$ $\approx$ -1.6(log T) - 1.6) up to the final day of
observations where flux densities remained above pre-flare levels:
S$_{5.5}$ $\approx$ 1.5 mJy and S$_{9}$ $\approx$ 1.0 mJy. MAXI X-ray
measurements (also shown in Figure 1) indicate a flare occurred at, or
just before, the time of the radio event, with statistical errors and
the lack of a radio peak preventing us from comparing the events to
any greater degree.

There is little or no evidence of a turnover in flux densities either
side of the flare, suggesting the peak of the event occurred towards
the middle of the gap between observations (i.e. $\sim$ MJD 55206.5)
at both frequencies. Our event appears to differ
from previous flares (e.g. 2000 Oct 20/21
and 2002 Dec 04/05; Figure 6 of Tudose \etal 2008) in a variety of ways. Firstly, our event's `peak'
(estimated to last from initial rise to the levelling off observed the
next day) has a duration of almost a full day, compared to the 2000
Oct 20 flare light curve whose peak appears to last less than a third
of that time. Secondly, our flare's rise is significantly steeper than many observed in the past; $\sim$ $\times$6
increase in 2 hours at 5.5 GHz and $\sim$ $\times$10 at 9 GHz, compared to the rise observed in 2002 Dec 04 where levels only
doubled over half a day.
However, the light curve following the 2002 Dec event does reveal a
similar gradual decline post flare (the 2000 Oct 20 light curve might
also show the start of a similar decay trend, but is cut short). Such
comparisons are difficult to justify, as Circinus X-1 was consistently
far brighter in the former epochs than that during which our
observations were carried out. As a result, we can only crudely
estimate the peak of the flare to have a flux density between 0.05 and
0.1 Jy at both frequencies, based on extrapolations of the initial
rise and decay rates.

The lower sections of Figures 1 and 2 show the spectral index (S$_{\nu} \propto \nu^{\alpha}$) of Cir X-1's core, based on the 5.5 GHz and 9 GHz flux densities. Indices remain negative for most of the pre-flare period, with an average value of $\alpha$ $\sim$ $-$ 1.0 but with significant scatter ($\sigma$ = 0.45) and statistical errors. This is, as expected, indicative of optically thin synchrotron emission, and consistent with previous measurements of Cir X-1's spectral index (e.g. Fig 10, Tudose \etal 2008). There does appear to be a rise to positive indices near MJD 55198, which would correspond to phase $\sim$ 0.5 (based on current ephemeris estimates). However, the day's data (2010 Jan 01/02) to which the points correspond, was particularly difficult to calibrate, making the flux densities unreliable. This effect might initially suggest an apastron radio flare (Fender 1997, Tudose \etal 2008), but it is more likely that the positive index is a result of errors and not a real physical effect.

\begin{figure}
\centerline{\includegraphics[width=3.4in]{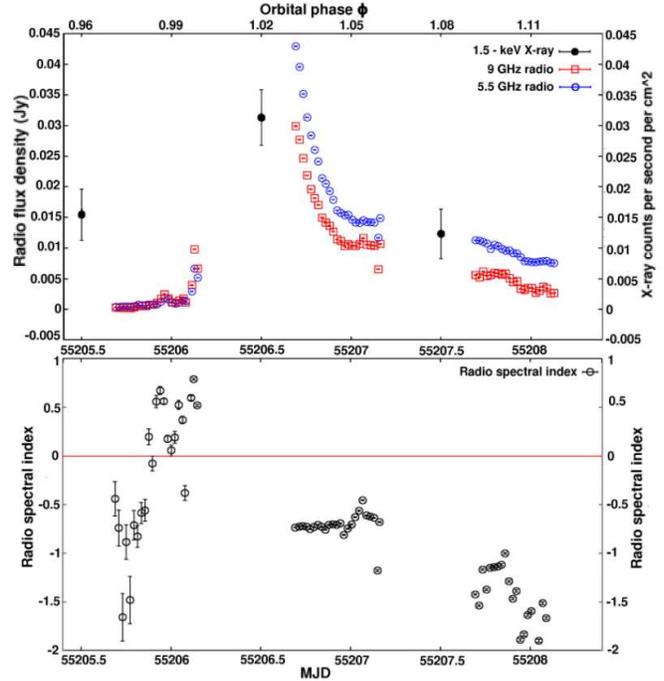}}
\caption{MAXI X-ray, ATCA-CABB radio (top panel) and radio spectral index (bottom panel) light curves for the flare event. Features such as the pre-flare `hump' (MJD $\sim$ 55205.95) are more easily visible.}
\end{figure} 

The spectral index begins a clear rise on the day of the flare,
reaching $\alpha$ $\sim$ 0.5 twice during the days observations,
firstly during times concordant with the `bump' 2 hours prior to the
flare, and then towards the end of observations with the rise of the
flare itself. By the following day, during flare decline, spectral
index has returned to negative values, $\alpha$= $-$ 0.7$\pm$0.1, with
very little scatter. Subsequent days show increasing levels of error
and scatter as values decline.

The behaviour of the light curves and radio spectrum are consistent
with that expected from either internal shocks created within outflows
as a result of quasi-continuous ejections (i.e. jet flows with higher
velocities during flares impacting with older slower upstream flows:
Kaiser, Sunyaev and Spruit 2000) or an adiabatic expansion of ejected
clouds of relativistic particles (as in van der Laan 1966), as is the
case for nearly all outbursts from XRBs. If we were dealing with such expanding knots of emission, one could expect these objects to eventually move a sufficient distance from the core to be resolved separately, which we do not observe. Such objects might still exist as part of the core emission, but if their velocities are far lower than those of the flows discussed in Fender et al. 2004, then there may not be sufficient time after outburst to directly observe the separation before the knots fade below detectable levels. Crude estimations of the velocities possible for this scenario can be calculated based on image synthesised beam size ($\sim$ 1.5'' for a 12 hour 9GHz CABB image) and the total time taken for core flux densities to stabilise after the flare ($\sim$ 6 days). Taking an assumed jet inclination of $\theta$ $\sim$ 5$^{\circ}$, and half the beam size as a required minimum distance for detectable change in the core structure, we find that such objects could remain unresolved if moving with Lorentz factors that do not exceed $\Gamma$ $\sim$ 5. This does not eliminate the possibility of ultra-relativistic flows re-energising media further downstream, since internal shocks can fade only to reignite during additional collisions (e.g. XTE J1748-288: Hjellming et al. 1998).

As mentioned earlier, flux
densities actually appear to level off towards the end of Jan 10
before commencing the gentler decay. There have been several examples
of multi-peaked flares from Cir X-1 in the past (Thomas \etal 1978,
Tudose \etal 2008) with varying numbers of peaks and timescales, but the gaps
in our observations prevent us from confirming such behaviour.

There is an unfortunate effect visible in the light curve, which
becomes more apparent in the spectral index plot, and that is a trend
towards more negative $\alpha$ (and lower flux densities) towards the
beginning and end of each run, particularly when the source is
brighter (i.e. post-flare); this is most easily seen in the
observations around MJD 55209. The effect is likely a result of Cir
X-1's low elevation (25$^{\circ}$ at minimum) during early and late
hours of individual observation runs (hence why it becomes more
evident once we switch from $\sim$ 9 hour runs to full 12 hour runs
after MJD 55200) leading to increased atmospheric opacities, or possibly, rising sensitivity to large scale structure as projected baselines decreased in length. While self calibration does appear to eliminate
this turn over effect (standard calibration methods fail to do so), it also alters source flux density in such a
way that makes measurements unreliable. Thus by using standard
calibration routines we retain flux density accuracy, while the
general trend and most important features of the light curve remain
clear.

\subsection{Imaging, modelling and subtractions}
\subsubsection{Original images}

Images were created for each day's individual data set, at both
frequencies. These are presented, for reference, in Appendix A (5.5
GHz in figure A1, 9 GHz in A2). The images were self-calibrated in an
effort to improve image fidelity and reduce the effects of artefacts
caused by the large flux density variations following flare onset. All
flux density measurements on the other hand (Table 1), are taken from
images that have not undergone self-calibration, but only the standard
calibration methods detailed in the MIRIAD guide. All {\em uv} data
with {\em uv} distances shorter than 6 k$\lambda$ at 5.5 GHz and 10
k$\lambda$ at 9 GHz were ignored in order to eliminate the more
diffuse emission from the surrounding nebula (see section 3.3). Note,
that we will first address the `normal' images from data sets before,
and several days after the flare, in which conditions are more stable
and similar from day to day, prior to discussing the images during and
shortly after the event. The stability during the pre-flare days also
allows us to combine the data from the first nine observations to
produce a deep 5.5 GHz map of the system's core and nearby region;
this image is presented in the lower left hand panel of Figure 3. Fig
3 also presents the normal robust weighted images for the four days
after the flaring episode (corresponding to images A13--A16 in
Appendix A1), as well as images in which the (variable) core has been
subtracted (see section 3.3) in an attempt to measure weak variability
on arcsec scales.

\begin{figure*}
\centerline{\includegraphics[width=7in]{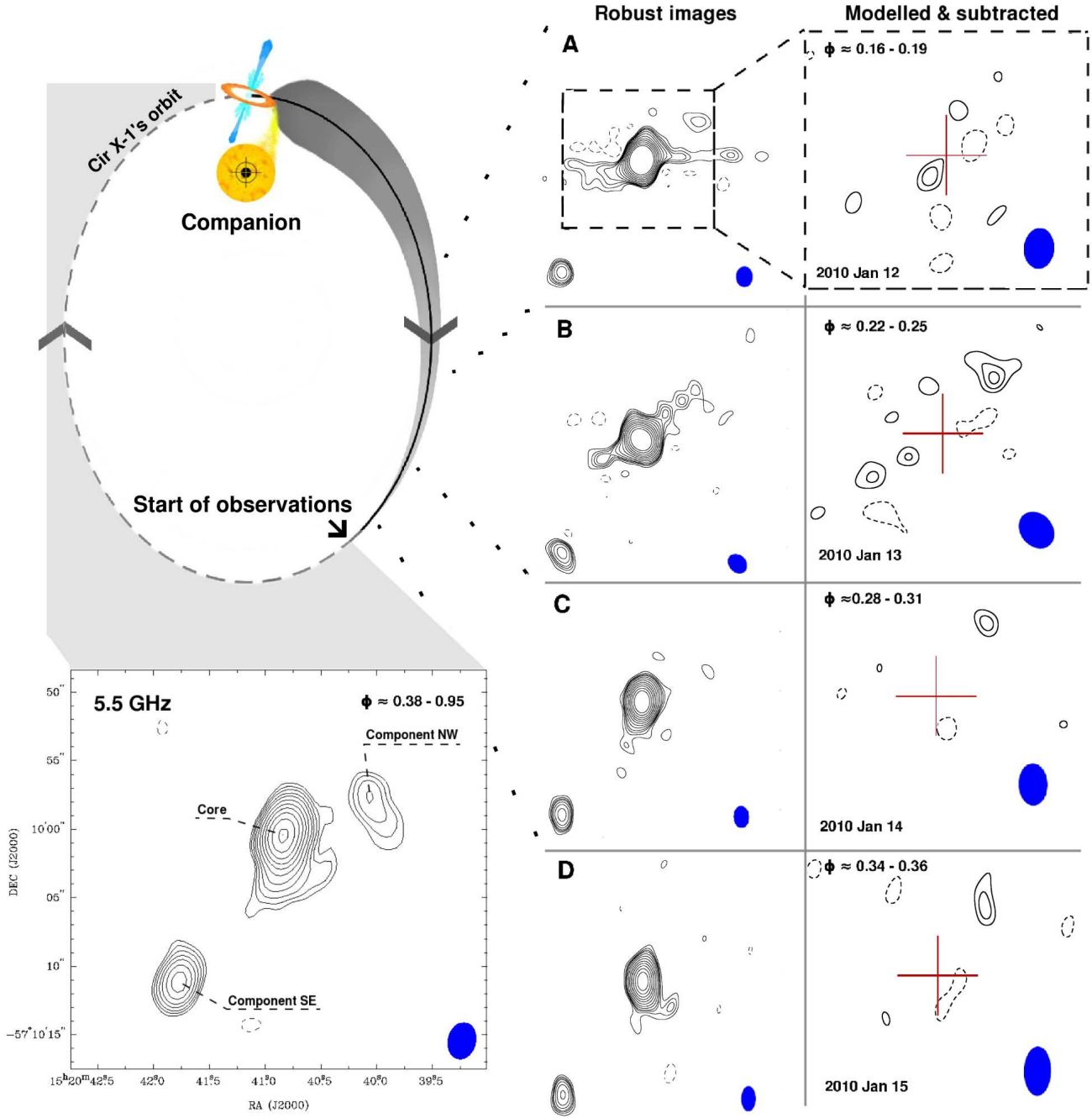}}
\caption{Illustration of Circinus X-1's orbital behaviour with related
  radio maps. Our intention is to show how the observed radio
  structure is related to the orbital motion of the system. The upper
  left diagram is a representation of the system, assuming a high mass
  companion and an eccentric orbit. The lower left image is a deep 5.5
  GHz radio map (beam size = 2.5 $\times$ 1.8 arcsec$^{2}$) of the
  core and nearby region of Cir X-1, combining data from the first 9
  observations, i.e. those taken prior to the flare event (ignoring
  {\em uv} data with {\em uv} distances shorter than 6 k$\lambda$),
  resulting in rms noise of 3.5 $\mu$Jy beam$^{-1}$. The images to the
  right show the unaltered data radio maps (left) and the final model
  subtracted maps (right) for the final 4 days of observations, after
  the flare event. Note that the right hand maps cover a smaller
  region around the core (as illustrated in the top set) in order to
  improve visible details. The core and nearby bright components
  referred to in section 3.2 are labelled. Contour lines in all maps
  are at -2.8, 2.8, 4, 5.6, 8, 11, 16, 23, 32, 45, 64, 90$\times$
  r.m.s. noise of each epoch (listed in related panels from appendix Figures A1 and C1). }
\end{figure*} 

The system's core (centred) is well defined in all images, and in
many of the 5.5 GHz examples shows a distinct south/south-east
extension (particularly evident in Figure 3's deep map) which can be
interpreted as evidence for an approaching jet, i.e. the jet component
aimed along our line of sight. The position angle of this jet is
similar, but appears to be slightly more southerly than that observed
in previous data sets (Tudose \etal 2008). Some of the individual
images also show a slight north/north-west protrusion, or general
elongation of the core along the north-south axis which cannot be
fully explained by beam shape alone, indicating the presence of the
(likely) receding jet. The maps produced from 9 GHz data show similar
elongation of the core, as well as extensions related to the
jets. However, unlike at 5.5 GHz, the 9 GHz northern jet structure
appears to be just as visible as its southern counterpart, and often
more easily so.

In figure 4 (upper row) we compare our deep pre-flare map to a single
day's observations from the pre-flare period of 2002 Dec. The
difference in the structure near the core is significant, as we can
clearly see a second bright component a few arc-seconds to the south
west of the core in the 2002 map which does not appear in our 2009/2010 map. This component
was that which varied following flare events, leading to the high jet
velocity estimates by Fender \etal (2004), and though not persistent,
structure has appeared in or very near this location multiple times in
the past at various levels of intensity (see Tudose \etal 2008).

The lower row of Figure 4 shows point source subtraction results based
on fits to the location of the core. It should be noted that the two
far components (SE and NW) in our maps would be less likely to appear
in the older map due to the lower signal to noise available (assuming
similar flux densities). The differences in residual structure around
the core is now more prominent. Firstly, though in both cases we
observe residual structure on opposite sides of the core void
(i.e. components likely related to the jet pair), we can clearly see
the south-eastern component is significantly brighter than its
north-western counterpart in the 2002 image, unlike our image which
shows more similar intensity between the two (southern residual peak
flux density of 79 $\pm$ 4 $\mu$Jy, northern residual with 75 $\pm$
4 $\mu$Jy). Secondly, the axis along which these structures lie is
different in the two images, with our map showing an axis ($\theta$ $\approx$ 170 $\pm$ 15$^{\circ}$) far closer
to north-south than 2002's near east-west orientation.
 
The images created from data sets during, and shortly after, the flare
(panels A10 - A13) show the artefacts caused by rapid variation in
brightness of a source during an observation (especially with an E-W
array with its instantaneous one-dimensional projection in the uv
plane), and as such much of what we see is not real, or requires very
careful interpretation.  There do appear to be resolved structures
along the SE-NW axis (i.e. the jet axis) that do not make up part of
the general elongation of the core, and thus we may be tempted to
believe they are jet related, however, given the time since flare
onset and angular distance of these structures ($\sim$ 5'') we can
quickly disregard these as real new components within the system. The
intensity of distortions is reduced as flux densities and decay rates
decline. The core at no point appears to resolve into two separate
bright components as has been observed in the past (i.e. as in Figure
4; Fender \etal 2004, Tudose \etal 2008).

Other than the core, the most notable structure in the images is the
separate emission component to the south east (labelled component SE
in Figure 3's deep map), located at approximately 15:20:41.75,
-57:10:11 (J2000). Upon initial inspection we would assume this to be a
possible component of the approaching jet, due to its proximity to Cir
X-1 and the fact its position angle would fit well with that of
previously observed outflow structures. The source is detected but
unresolved at both frequencies, with an average spectral index of
$\alpha$ $\sim$ -1.4. The angular distance from the core to this
component is $\sim$ 13'', which, based on previously estimated outflow
Lorentz factors ($\Gamma$ $>$ 10, distance $\approx$ 6.5 kpc), would
mean flux density variations linked to core outbursts and subsequent
re-energisation via the jet will be delayed by $>$ 30 days (assuming
constant flow velocity). We may still be able to observe re-energisation of the component by flows from an older core outburst, with a large phase offset resulting
from its distance. Light curves for this component at both
frequencies are visible in Figure 5, with errors based on noise levels
of each image. Measurements from 2010 Jan 9, 10 and 11 (A10--A12) are
not included as low-level artefacts were noted to extend from the core
and overlap with the component emission, reducing the validity of
those estimates. There is at best only a hint of variation in the flux
density of this component at 5.5 GHz, but some evidence at the $\sim
2\sigma$ level at 9 GHz.

A second fainter component also regularly appears north-west of the
core in the 5.5 GHz images (15:20:40.1, -57:09:58 [J2000] - labelled component
NW in Figure 3), though remaining undetected at 9 GHz. The source
appears resolved in both the deep map and individual epochs, often
displaying a south western extension. Again, we are dealing with an
angular distance ($\sim$ 7'') that would likely involve long delays
between outbursts and subsequent reactions in the component ($\sim$ 20
days). The 5.5 GHz light curve is included at the bottom of Figure
5. As with component SE, there is no unambiguous evidence for
variability.

\begin{figure}
\centerline{\includegraphics[width=3.4in]{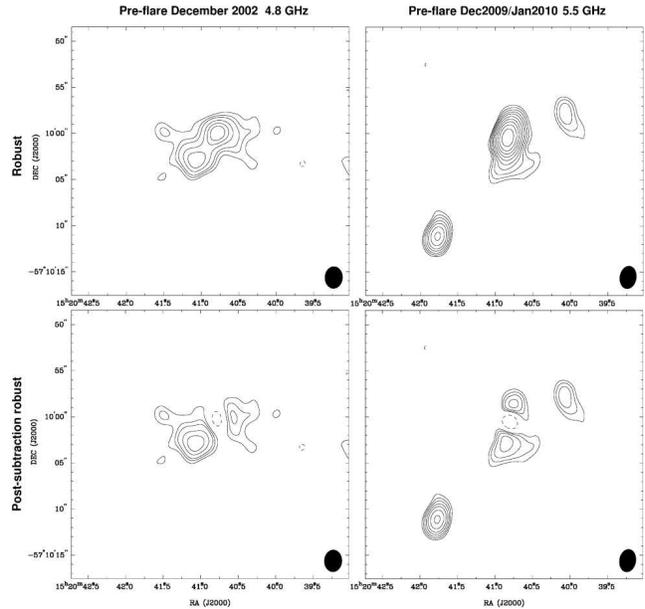}}
\caption{Point source subtraction maps. The left hand column shows
  contours (-2.8, 2.8, 4, 5.6, 8, 11$\times$ r.m.s. noise of 55
  $\mu$Jy beam$^{-1}$, beam size = 2.4 $\times$ 1.9 arcsec$^{2}$) for
  data from pre-flare observations taken on 2002 Dec 02 (See Fender
  \etal 2004). The right hand column are contour images made using the
  nine pre-flare observations from our data-set (-2.8, 2.8, 4, 5.6, 8,
  11, 16, 23, 32, 45, 64, 90$\times$ r.m.s. noise of 3.5 $\mu$Jy
  beam$^{-1}$,beam size = 2.5 $\times$ 1.8 arcsec$^{2}$) The upper row
  shows un-altered images using a weighting scheme of robust =
  0.5. The bottom row show the same images but with a point source
  fitted and subtracted from the approximate location of the system's
  core. Though there are clear differences in structure near the core
  itself, it should be stressed that the lowest contour levels in the
  2002 images are close to the flux densities measured from the peak
  of component SE; i.e. there is a good chance neither component SE
  nor NW would be visible in the older map assuming their flux
  densities remain constant.}
\end{figure} 

In an effort to determine the origin of these emission sources, we
reviewed past observations of Cir X-1. Reviewing images from Tudose
\etal 2008, we found several with distinct compact emission near the
location of component SE (a total of 5 epochs). Data sets from
individual days of pre-CABB observations tend to yield images with
noise levels close to the flux of these components, making detection
difficult. Therefore, we created stacked data sets which combined
observations of close epochs (i.e. within a maximum of 2 orbital
periods difference in observation time) to produce deeper maps in an
effort to track down further evidence of the components. These were:
July 1996, October 1998, May 2001 and December 2002 (See Tudose \etal
2008 for details on these data sets). Component NW remained undetected
in all revised images; however, 3$\sigma$ $>$ 0.05 mJy in all cases,
which based on our flux measurements would make the component
difficult to detect. Component SE was visible in nearly all
stacked images with the following flux densities: Dec 2002 - 0.17
$\pm$ 0.02 mJy; May 2001 - 0.25 $\pm$ 0.03 mJy; and Oct 1998 - 0.20
$\pm$ 0.04 mJy. The exception was Jul 1996, where we had difficulty
distinguishing separate components due to the abundance of structure
near that portion of the image. These fluxes are not dissimilar from
the range measured in our data.

Linear polarisation was detected ($>$ 3 $\sigma$ levels) at the
position of the core on 10 Jan 2010 at 9 GHz. No other images show any
distinguishable polarisation in regions of interest and the single
detection is relatively weak ($<$ 2\% of core flux density: $<$ 0.1
mJy), thus few strong conclusions can be garnered from it. Based on estimates quoted in Tudose \etal (2008) a correction of $|\Delta\theta|$ $\leq$ 9$^{\circ}$ must be applied to account for Galactic Faraday rotation. Vector
orientation indicates
electric fields aligned away from the jet axis, but not entirely
perpendicular; similar to what was observed from the core in previous
epochs (see section 7 of Tudose \etal 2008).

\subsubsection{Difference images}

\begin{figure}
\centerline{\includegraphics[width=3.4in]{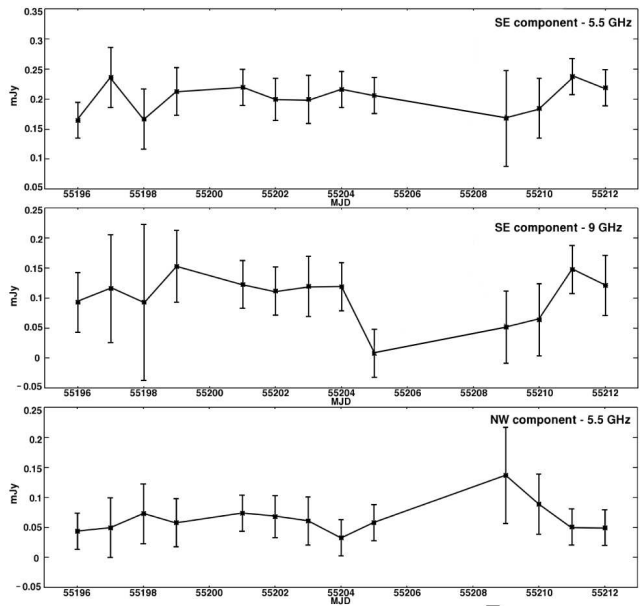}}
\caption{Radio light curves for components SE and NW (see Figure 3,
  bottom left). Values are from peak flux density measurements. The
  large gap in points is a result of unreliability of measurements
  from data taken during the flare and its decay.}
\end{figure}

In order to compare data sets on a day to day basis in search of
structural changes associated with Cir X-1's flare, it was prudent to
select a single day's data to act as a reference. This observation
could then be subtracted at the in the uv plane (prior to construction
of images) from each epoch, leaving only those regions of flux density
which had changed in intensity. The reference data set would have to
fall on a day sufficiently distant in time from a flare event, so that
any subsequent variation in flux levels had diminished, both of the
core and any nearby structures that may have reacted to the
event. Based on the flare event captured in our observations, we can
see that even 6 days after flare onset, the core continues to remain
above pre-flare flux density levels. It is best to select a day that
falls beyond a half-period after a flare event as a reference, i.e. 8
days or more. In case of a flare having occurred prior to that in our
data, the reference would fall on 2010 Jan 01. Unfortunately we must
also take into account data quality, as some days include gaps (or
shorter observation durations) and increased levels of RFI. As
mentioned in section 3.1, Jan 01/02 data suffered from problems
making it an inadequate reference choice. Data from 2010 Jan 04
suffered from neither gaps, unexpected errors, nor high levels of RFI,
and additionally covered a near full 12 hours, meaning it could not
only be cropped to match hour angles of the earlier half of
observations (each $<$ 9 hours long) but would also cover most of
subsequent observations for effective comparisons.

Appendix B (Figures B1 and B2) shows difference maps for all days at
5.5 GHz and 9 GHz respectively. The days prior to Jan 04 show excess
flux at, or near the core; whereas the days following, but prior to
the flare event, show regions of over-subtraction in the same
location. This is a logical result from the trend in fluxes listed in
Table 1 and would suggest a gradual decay in flux density over the
first 10 observations, perhaps following a flare event preceding that
of Jan 09. There are a pair of structures visible in the December 5.5
GHz images (panels C1 and C2) towards the north-west which should be noted, as they appear along the known jet axis and in the same
locations in each of the images, then go undetected in most pre-flare
maps. Images from the days following the flare event remain heavily
affected by artefacts; however, the asymmetrical nature of some of the
structures (e.g. C13) suggest that there may be some real emission
towards the south-east which may persist over several days. A similar
argument can be made for the structures in panel C14, and the emission
component that is visible to the north-west persists in panels C15 and
C16. Furthermore, its position is concordant with the structure
nearest the core in panels C1 and C2. The left hand panel of Figure 6 shows the results of stacking data used for panels C1, C2, C15 and C16, in which one can clearly see the northern structure nearer the core. Though not shown in the image, the farther north-western structure from panels C1 and C2 was also visible, though only at a 3 $\sigma$ level. The extension visible to the north-east of the core appears to be caused by an streak artefact of unknown origin running through the core perpendicular to the jet axis (further effects of this artefact can be seen to the core's south-west).

\subsubsection{Model subtracted images}

It is clear that the flare artefacts severely hinder our ability to
interpret the images after Jan 09, with core variation continuing and
causing elongation along an axis with an unfortunate similarity to
that of the jet. Thus a second level of subtraction would be required
to reveal structural variation in such images. This involves modelling
the flaring core, based on the variation observed from measured light
curves of each observation.

\begin{figure}
\centerline{\includegraphics[width=3.4in]{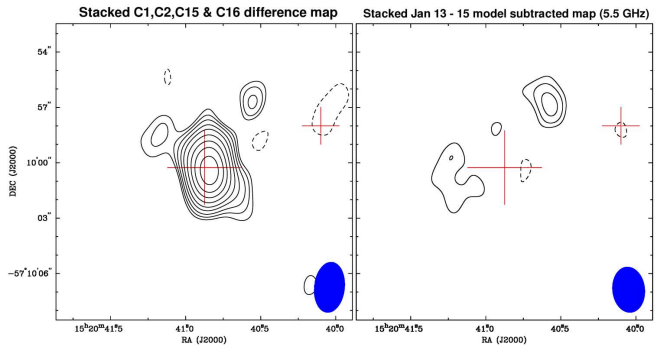}}
\caption{Stacked difference and model subtracted image maps. The left hand image combines the difference data from 2009 Dec 30, Dec 31, 2010 Jan 14 and Jan 15 (panels C1, C2, C15 and C16 from appendix Figure B1: folded orbital phase $\phi$ = 0.28 - 0.46, beam size = 2.75 $\times$ 1.66 arcsec$^{2}$), the right hand image combines model subtracted data from 2010 Jan 13, 14 and 15 (orbital phase $\phi$ = 0.22 - 0.36, beam size = 2.48 $\times$ 1.76 arcsec$^{2}$). Both images use contour levels of -2.8, 2.8, 4, 5.6, 8, 11, 16, 23, 32, 45, 64, 90 times r.m.s. of 13 $\mu$Jy. The larger cross marks the core, the smaller the position of component NW. An additional (i.e. unrelated to component NW) north-western structure can be seen in both images.}
\end{figure} 

The process is as follows: each day's difference data set (i.e. with
the Jan 04 reference data already removed) is divided into smaller
sections, approximately 5 minutes in length ($>$100 data segments per
observation). These individual sections are then converted into images
and a measurement of the maximum flux density is taken from a 30
arcsec$^{2}$ region centred on the core. It should be noted that no
effort to clean the image is made prior to the measurement, as it
results in no improvement to image fidelity for such small data
chunks. Though these snapshot images are of very poor quality, the
light curves that result from this process are very similar to that of
Figure 1 suggesting relatively accurate representation of behaviour at
the core. The flux density measurement of each 5 minute segment is
then used to construct a 5 minute data model, representing a single
source of the same flux density at the position (based on measurements
made from the unaltered data) of Cir X-1's core. Using the MIRIAD task
UVSUB, one can then create new 5 minute data segments with the models
subtracted, and recombine them into a complete observation
(UVCAT). This final data set can then be inverted and cleaned as
normal, eliminating as effectively as possible the recorded core behaviour.

The results of this process on the post flare images are shown in
Appendix C. Unsurprisingly, the subtractions were least successful on
the Jan 10 images owing to the rapid rates of variation soon after the
flare. Many of the structures around the location of the core are likely
to be remnants of flare artefacts or the result of phase errors, inferred from the symmetrical
or anti-symmetrical layout of strong positive and negative counterparts (easily visible in
the 5.5 GHz Jan 10 image). Though the Jan 11 images are cleaner by
comparison, we must remain wary as we are still dealing with a day of
high flux densities and decay rates. Most visible structures can be
disregarded, owing to their numbers, negative counterparts, and almost
random arrangement.

The final four model subtracted images are also shown in Figure 3 (far
right column) together with their unaltered counterparts (described in
section 3.2.1), in an effort to summarise the behaviour that occurs
within Cir X-1. Jan 12 images (row A of Fig 3) show some emission to
the south-east (i.e. along the jet axis) of similar size and position
at both frequencies, leading to jet velocity estimates with $\Gamma$
$>$ 10 (assuming a relation with the flare of Jan 9th). However, we also observe a negative component at 5.5 GHz whose
position would suggest these may be an artefact pair. We also observe
strong emission to the north at 9 GHz, but the lack of a 5.5 GHz
counterpart means we cannot confirm if it is real. The images from Jan
13 onwards show the strongest evidence of real structural variation
beyond the core. These were the images that showed the lowest level of
artefacts caused by the flare, prior to core subtraction, and thus are
the most likely to display real emission structures rather than
residual artefacts. 

This leaves three components of interest in the 5.5 GHz images, two of
which have possible counterparts at 9 GHz that do not appear to be
related to side-lobes. The component nearest the core at 5.5 GHz on Jan
13 may be related to the negative component on the opposite side of
the core, and without a confirmed counterpart at 9 GHz is best
ignored. The remaining two components reside on opposite sides of the core and have possible 9 GHz counterparts appearing a
significant distance further from the core (1-2''), yet they remain
along a very similar axis. The northern component at 15:20:40.56 -57:09:56.7 (J2000) does appear to have a negative counterpart at a similar distance to the south-east of the core, and though it is slightly smaller and fainter, may still indicate effects of flare artefacts. However, support for the reality of this northern structure is that it persists at 5.5 GHz through Jan 14 and 15 images, and the
position also coincides with that of the near-core northern component
observed in the Dec 30 and 31 difference maps (see section
3.2.2). Both positive structures are not seen again at 9 GHz, nor does any
significant new structure arise over the final 2 days. The result of stacking these final three days of model subtracted data is shown in the right panel of Figure 6, where we can see strong evidence of the northern component ($>$ 5$\sigma$) and weaker signs of south-eastern residual structure ($\sim$ 3$\sigma$).

Flux density
variation of the northern component over the 3 days is well within the noise
levels of the individual images. The emission lies approximately 3 -
5'' away from the core, and appeared $\sim$ 3 days after the peak of
the flare event. If we assume it is real, then it is unlikely to be
related to our observed flare event, as that would require Lorentz
factors of $\Gamma$ $>$ 35. However, if we consider the possibility of
an earlier flare (16.6$\times$n$+$4 days delay) being the source of re-energisation, then the calculated velocities may be significantly reduced (e.g. $\sim$ 20
day delay gives $\Gamma$ to be 6 - 10).

As a test of our methods we investigated whether the process of
removing the generated model from the data could inadvertently create
solitary structures such as those in the images. We created model data
with a single point source at Cir X-1's location (much like the
subtraction model) but included noise within the flux density values
with a combination of sinusoidal variation and random number
components. The resulting models strongly resembled a single day's
high resolution light curve of Cir X-1's core (with flux density
levels set to be similar to those seen during flare decay). We
proceeded to image this model data and run it through the same light
curve extraction and model subtraction routines we used on the real
data (with varying interval sizes). The result was a near complete
removal of the source each time (a weak `core' would remain with flux
density $<$ 1\% of original levels), with flux densities of residual
surrounding structure in the image reaching only a few $\mu$Jy
(i.e. $<$ r.m.s. noise levels in standard images) and distributed in a
spoke like manner rather than clustering.

\subsection{Large scale structure}

\begin{figure*}
\centerline{\includegraphics[width=7in]{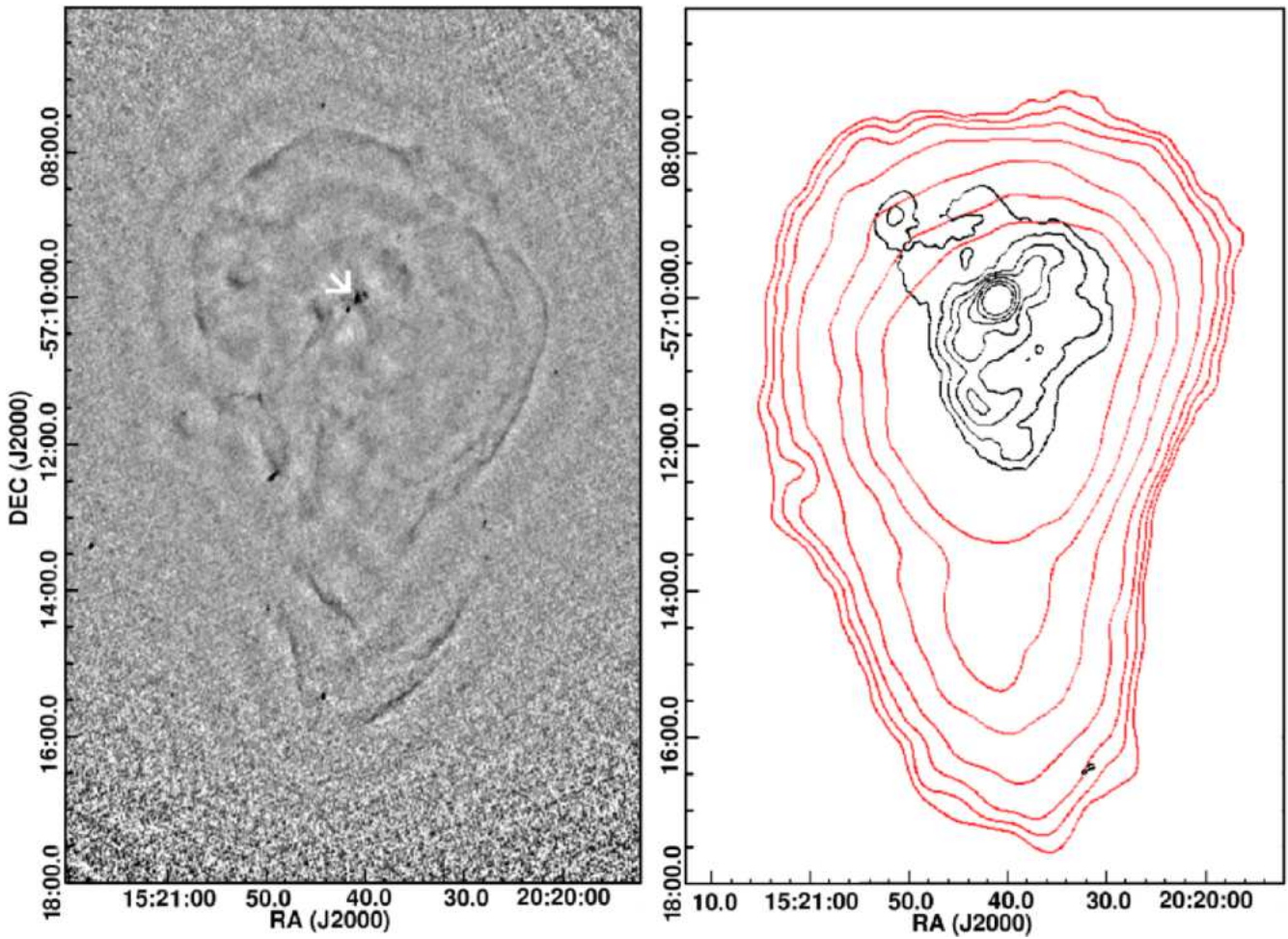}}
\caption{Large scale radio images of Circinus X-1's jet powered nebula. The left hand 5.5 GHz radio map (beam size = 2.3 $\times$ 1.6 arcsec$^{2}$) was produced using combined data from the first 9 days of observations (i.e. the same data as the bottom left panel of Figure 3, but using complete {\em uv} distance coverage) and shows the detailed filament like structures outlining the edges of the nebula, with Cir X-1's core marked by the white arrow. The increase in noise towards the bottom of the image is a result of primary beam correction - i.e. approaching the beam perimeter. To the right we show a contour map of the same region using 1.4 GHz data from Aug 2001 (Black, contour levels of 1, 1.4, 2, 2.8, 4, 5.6 and 8 times the rms noise of 2 mJy beam$^{-1}$) and Sep 2004 (Red/Grey, contour levels of 1, 4, 8, 16, 32, 48 and 64 times the rms noise of 0.7 mJy beam$^{-1}$), the difference between the two being due to array configuration (i.e. shorter baselines were available during the Sep 2004 observations). This is the same data use to produce figures 2 (middle) and 3 in Tudose \etal 2006, and illustrates how much of the large scale structure appears beyond the region outlined by the filaments in our maps.}
\end{figure*}

By combining the observations from the first 9 days we were able to create a deep 5.5 GHz radio map of Circinus X-1 and its surrounding region (Figure 7, left), with rms levels in the majority of the map not rising above 3.5 $\mu$Jy beam$^{-1}$. These low noise levels allow us to see detailed filament-like structures outlining the known layout of the Cir X-1's jet powered nebula, and what appear to be faint (compared to the core) less diffuse regions within the larger nebula structure. We compare our map with contours created from the data used in Tudose \etal 2006 (Figure 7, right). The red/grey contours are those from 1.4 GHz observations with the ATCA in a short baseline configuration (EW214: minimum baseline of 31m, maximum of 4500m) taken in September 2004, showing the outer regions of the nebula surrounding Cir X-1. The black ATCA contours (configuration 1.5A: minimum baseline of 153m, maximum of 4469m) are from August 2001 and are also at 1.4 GHz, but the observations did not include baselines as short of those from Sep 2004, and thus we observe slightly more detailed structure closer to the core. The asymmetry of the structure can be observed on both these scales: an extension of the nebula towards the south. This has been mostly attributed to projection effects as a result of the approaching south-eastern jet, and receding north-western jet. However, there are also bends in the jets (evident in the black contours) which may be attributed to precession of the jet (much like SS433) or interactions with higher-density material along the jet's path which cause the flow to deviate. 

In our deep map we can see numerous filament structures outlining the nebula's edges, as well as multiple components along the jet axis, including the radio counterparts to the X-ray synchrotron `caps' reported in Sell \etal (2010). Though they form circular outlines, the filaments are unlikely to be the residuals of side lobe structure, based on comparison to the corresponding beam patterns; they are too large, intense and sharply defined (even prior to the application of cleaning algorithms). Much of the more diffuse structure within the confines of the filaments is affected by beam distortions, causing large scale `lumpiness' or corrugation, though several sharper intense regions stand out (such as the caps). The majority of the outlying filaments have peak flux densities of $\sim$ 2$\times$$10^{-5}$ Jy beam$^{-1}$, but since the 6A ATCA array configuration uses particularly long baselines (especially in comparison to those used in Tudose \etal 2006), it is likely much of the emission from these regions has been resolved out. This is also indicated via comparison of the flux levels with those measured at the same locations at lower frequencies, yielding an extremely steep spectrum ($\alpha$ $<$ $-$4). It should also be noted that though we have corrected for the response of the primary beam, the distance of the filaments from the pointing centre of the observation ($\sim$ 2.5 arcmin for the `ring' filaments, 4 - 5 arcmin for those to the south) may still add to the uncertainty in flux density measurements as a result of uncertainties in the primary beam model itself.

None of the filaments are detected at 9 GHz or in polarisation maps, and estimates for individual structures show no region has minimum brightness temperature T$_{b-min}$ $>$ 0.5$\times$10$^{4}$ K, meaning either synchrotron or bremsstrahlung processes can be the cause. Nonetheless, these filaments clearly define an outline similar in form to that of the nebula, including the southern extension, though smaller in comparison to the full low frequency emission (red contours). Assuming the validity of a bremsstrahlung scenario, a comparison may be made with the nebula of Cygnus X-1 where we observe the thermal bremsstrahlung emission from ionised gas produced behind the bow shock of a jet impacting on the ISM, for which temperature estimates give T $\sim$ 10$^{4}$ K (Gallo \etal 2005).

It is also possible that we are observing boundaries related to the synchrotron emission bubble formed by the outflows and being reheated by continuing jet emission pointed towards us. Tudose \etal 2006 reported flattening of the nebula's spectrum towards its edges, in particular in the north east and south west extremes: regions in our maps where we observe longer, unbroken emission filaments. The interpretation that followed was that the sites were regions of increased interaction between the ISM and accelerated particles. The filament like morphology may simply be a result of density fluctuations in the ISM, or perhaps, if the jet does precess or the large scale structure of the outflows has changed over time, they may be regions of jet-ISM interactions much like those observed in nebula W50 (Dubner \etal 1998). It could easily be imagined that, if this nebula resembles a layout much like that of SS433 and W50 but orientated towards us (inferred from the jet inclination: Fender \etal 2004), then, as suggested by the contour maps, the diffuse emission we see at lower frequencies would appear to extend beyond brighter zones reheated by the jets due to the `middle' of the nebula being wider than the tapered ends (through which we are viewing the system).

\section{Discussion}

Previous observations of Cir X-1 taken during its brighter past have shown most of the system's activity and structural variation occur south-east of the core, which is likely to be the direction of the approaching jet. Indeed there have been several epochs during which the emission of a distinct south-eastern component only $\sim$ 2 - 3 arc-seconds from the core has outshone the core itself (2001 May 29 - Fender \etal 2004, Tudose \etal 2008). There has also been some evidence of the receding jet (1996 July 02, 2000 October 25); however, the receding jet never appeared without activity from its counterpart, with the latter remaining dominant in comparison.

Initial inspection of images such as the bottom left panel of Figure 3 would suggest the current trend is similar, but with the core being significantly diminished. We observe a bright component (component SE) $\sim$ 13 arc-seconds to the south east of the core and a fainter counterpart to the north/ north west (component NW), slightly nearer but still over 10 arc-seconds away. However the nature of component SE can be easily scrutinised. Firstly, as well as being unresolved in the radio, it was also found by Sell \etal (2010) to be unresolved in X-rays and coincident with a bright IR point source. Subsequently, Sell \etal were unable to statistically rule out the possibility that the component is in fact a background source. If the emission was indeed related to the jet output from Cir X-1, we might expect some level of long term variation in flux density (to match Cir X-1's core decline over the past decades) or position (as a previous ejection event). Archival data not only revealed it to exist in the same position during several epochs over the past decade or more, but that its flux density has not varied significantly beyond the range of values measured from our observations, and has not followed any particular trend. With this additional information in mind it becomes easier to disregard component SE as jet related emission, but it will require further observation before a firm conclusion can be made.

The northern counterpart (component NW) remains a strong candidate for jet emission. The source is resolved in both radio and X-rays, as per Sell \etal 2010, and was not found to have an IR counterpart. It remains undetected in any past observation stacks (bearing in mind if at the same low flux density it would fall below the noise), though there has been evidence of intermittent structure near its location in individual epochs detailed in Tudose \etal (2008); specifically 2000 October 20/21 and 2001 May 27.

The strongest indication of structural variation in our maps
originates in regions to the north west of the core. The structure
visible in model subtracted images from Jan 13 through 15 (and
possibly Dec 30, 31) lies at an angle similar to that of the jet and
thus may be related to the receding outflow, though we observe
comparatively little change in the south. This behaviour goes against
what is expected of the system based on past observations, with the
greatest level of activity coming from a brighter approaching jet
located to the south. This difference could simply be due to intrinsic
asymmetry of interacting material around the binary, but jet
precession could be an alternative explanation. For a small angle to
the line of sight, as proposed, even a small degree of precession
could account for a reversal of orientation with the approaching jet
appearing in the north western quadrant. It should be noted that no
significant change in jet axis has been observed in radio images that
have been made of the core over the last decade and a half prior to
our observations (Tudose \etal 2008), suggesting that if any change in
orientation has occurred, it was sudden and recent. Nonetheless, a
near 180 degree reversal of the projection of the jet axis, though
plausible, remains suspicious, and suggests that the asymmetric flow
theory is more likely.

We may also be dealing with a kinked or deviated outflow. Though we
already see evidence of the south eastern outflow being curved on
arc-minute scales, there may be bends earlier in the outflow that
result in increased beaming effects along portions of the jet further
inclined along our line of sight. Such could be the case for component
SE, providing explanations for its stable position and compact
structure if it were related to a persistent jet kink. Again, if
dealing with a generally low jet inclination, significant deviations
of the outflow direction could cause emission activity within the
approaching jet to appear either side of core via projection along our
line of sight.

The point source subtraction images shown in Figure 4 also provide
evidence for precession, showing a possible axis that is almost a
north-south orientation. Additionally, based on the greater degree of
symmetry in residual structure around the core, our images suggest a
higher inclination for the jets than that observed in the past. This
latter conclusion might be seen to weaken the justification for using
low jet inclination as a possible explanation for the observed
behaviour in our images. However, just as observed in the outflows of
SS433, the inclination of a flow can vary as we move further from the
core, in this case as a result of either precession, flow deviation,
or a combination of both.

\section{Summary}

We have presented the results from ATCA-CABB radio observations of a
complete 17 day orbit of Circinus X-1 at 5.5 and 9 GHz. We
successfully captured the rise and decay of a periastron flare event
from the system's core and mapped the structure of the system on each
day during this period, during which it transpires Cir X-1 was in an
historically faint radio state. Prior to the flare event the system
was very stable at $\sim$ mJy flux densities. After the flare there is
some evidence for significant spatially resolved changes in the images
on arcsec scales in carefully modelled and core-subtracted images. The
strongest sign of structural variation appears north west of the core,
in the region previously associated with the direction of the receding
jet. We interpret this unexpected behaviour as either an indication of
asymmetry in the surrounding, or alternatively, of a change in outflow
direction via precession or curvature of the jets. A change in jet
orientation is also supported by comparisons of residual structure
around the core in our maps to that of maps made from observations
taken in 2002, showing an apparent `rotation' of the arcsec-scale
resolved core structure.

Deep radio maps of the area around the core show several persistent
sources likely associated with the jets, and others that are
counterparts to previously reported X-ray shocks. Large scale mapping
of Circinus X-1's surrounding region reveals filaments outlining the
known structure of the system's jet powered nebula, which, assuming a
low inclination for the system's jets (as has been previously
claimed), suggest the system's configuration may resemble that of
SS 433 and W50, but viewed longitudinally.

\section{Acknowledgements}
D.E.C. and J.W.B. acknowledge support from the United Kingdom Science and Technology Facilities Council. The Australia Telescope Compact Array is part of the
Australia Telescope which is funded by the Commonwealth of Australia
for operation as a National Facility managed by CSIRO. This research has made use of the MAXI data provided by RIKEN, JAXA and the MAXI team.

\clearpage
\appendix

\begin{minipage}{6.4in}
\section{Un-altered data radio maps}
The following maps (including those in subsequent appendices) use ATCA-CABB data that has undergone both normal calibration and phase calibration routines prior to final deconvolution. Maps were cleaned using multi-frequency clean subroutines.
\end{minipage}

\begin{figure}
\begin{minipage}{7in}
\centerline{\includegraphics[width=6.8in]{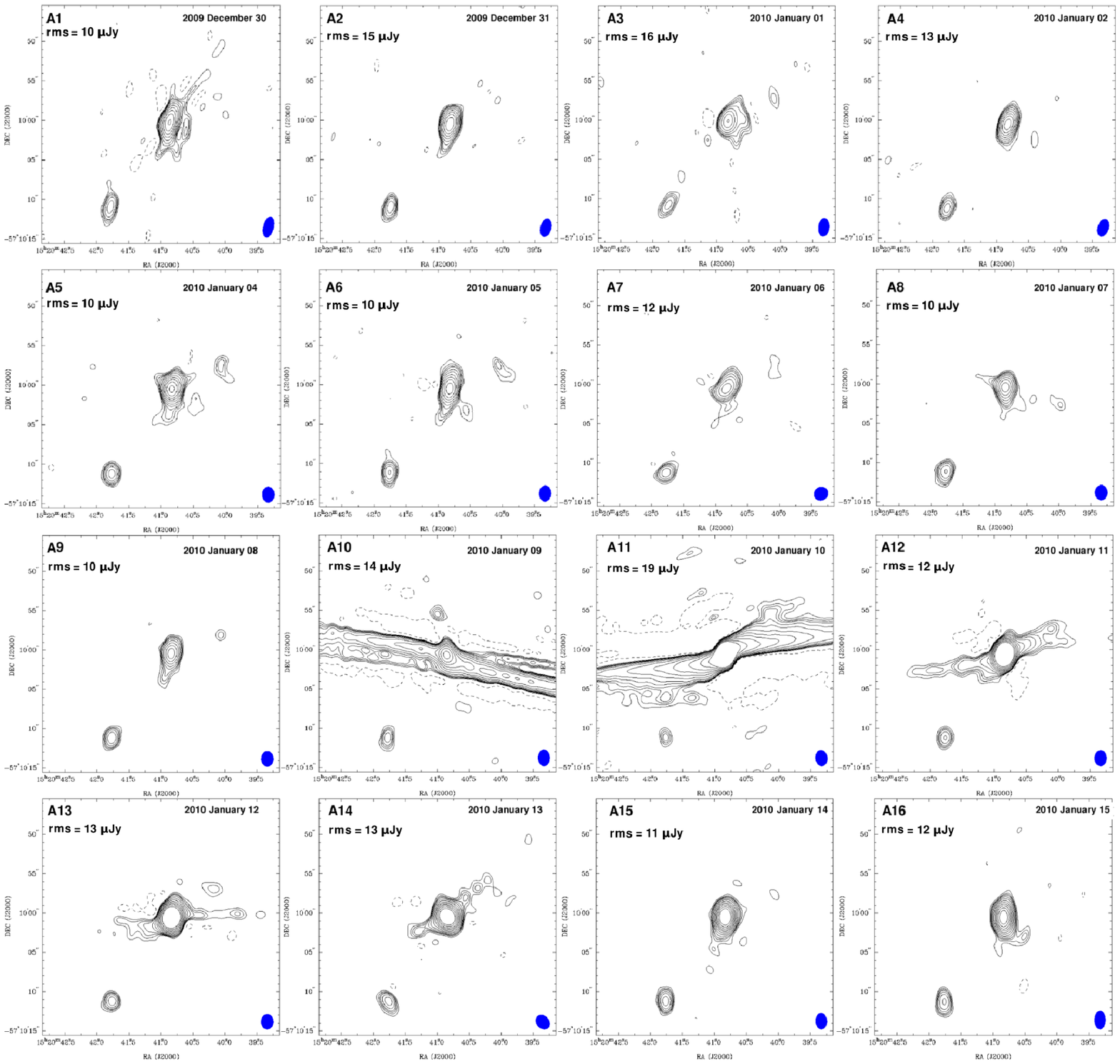}}
\caption{Radio maps of Circinus X-1 at 5.5 GHz. Weighting is determined with a robust factor of 0.5 (an optimal compromise between natural and uniform weighting). Contour lines are at -2.8, 2.8, 4, 5.6, 8, 11, 16, 23, 32, 45, 64, 90$\times$rms noise of each epoch (matching the scheme used in Tudose \etal 2008, rms listed in the top left corner of each panel). Beam sizes are approximately 4 arcsec$^{2}$. Artefacts caused by flare variability are easily visible in images A10 through A13.}
\end{minipage}
\end{figure} 

\clearpage

\begin{figure}
\begin{minipage}{7in}
\centerline{\includegraphics[width=6.8in]{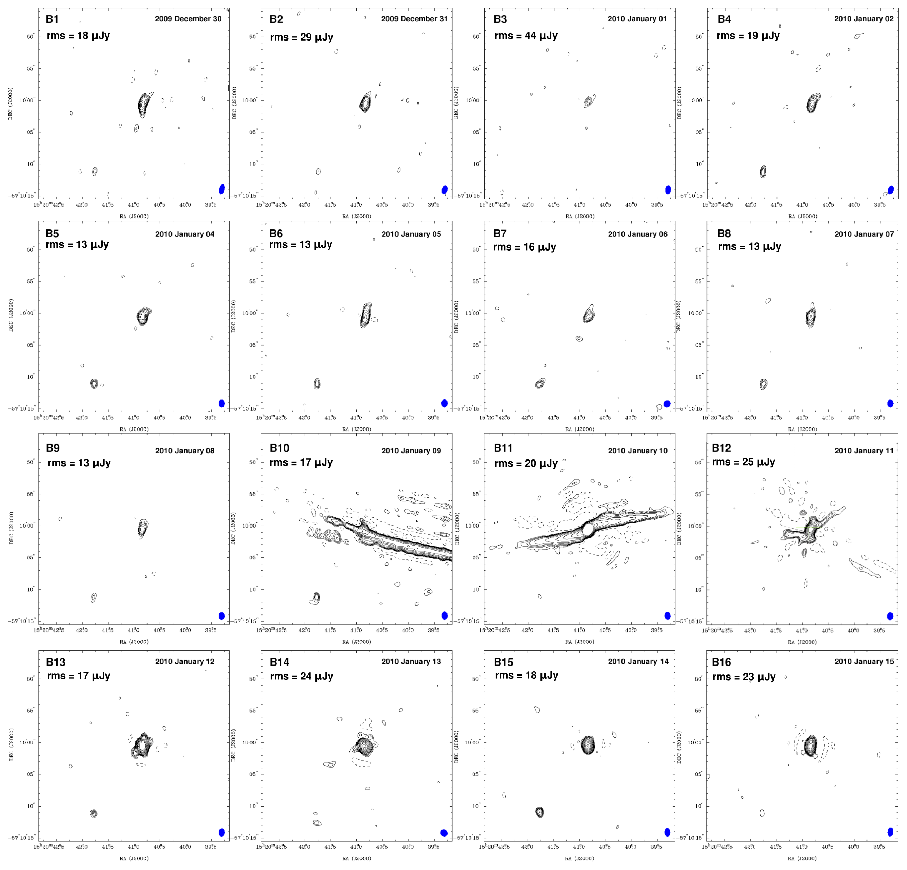}}
\caption{Radio maps of Circinus X-1 at 9 GHz. Weighting is determined with a robust factor of 0.5 (an optimal compromise between natural and uniform weighting). Contour lines are at -2.8, 2.8, 4, 5.6, 8, 11, 16, 23, 32, 45, 64, 90$\times$rms noise of each epoch (matching the scheme used in Tudose \etal 2008, rms listed in the top left corner of each panel). Beam sizes are approximately 1.5 arcsec$^{2}$. Artefacts caused by flare variability are easily visible in images B10 through A11.}
\end{minipage}
\end{figure}

\clearpage
\section{Difference radio maps}

\begin{figure}
\begin{minipage}{7in}
\centerline{\includegraphics[width=6.8in]{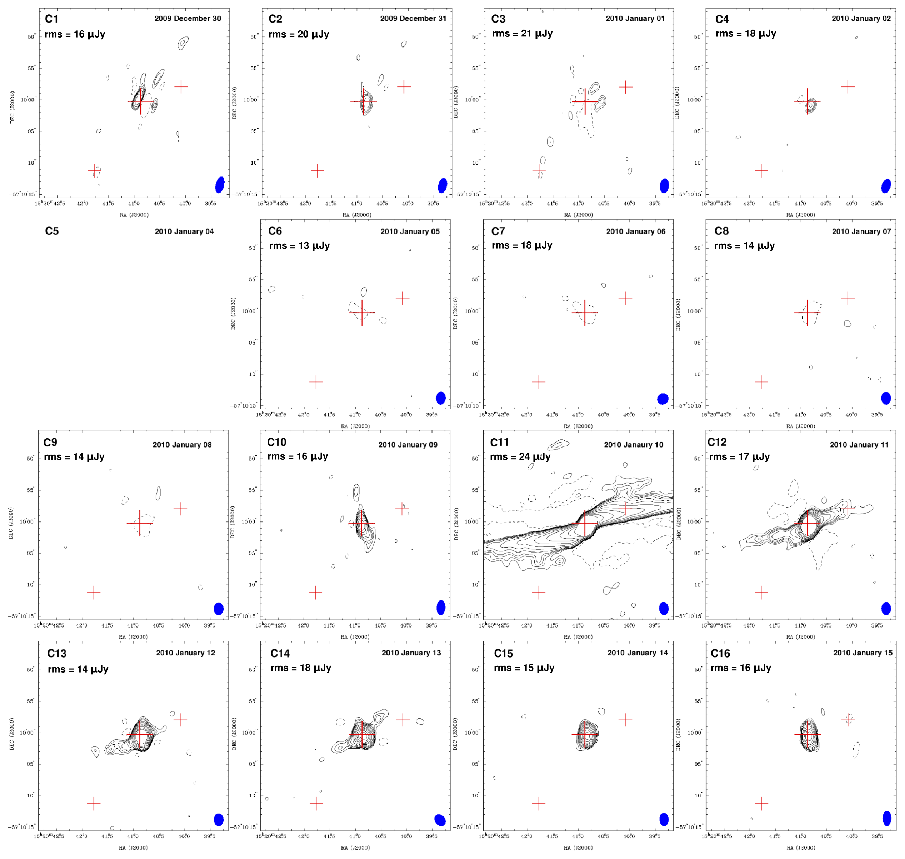}}
\caption{Difference radio maps of Circinus X-1 at 5.5 GHz. Each epoch has had data from 04 January 2010 subtracted prior to imaging (thus no image is available for 04 Jan itself). Weighting is determined with a robust factor of 0.5 (an optimal compromise between natural and uniform weighting). Contour lines are at -2.8, 2.8, 4, 5.6, 8, 11, 16, 23, 32, 45, 64, 90$\times$rms noise of each epoch (matching the scheme used in Tudose \etal 2008, rms listed in the top left corner of each panel). The larger cross marks the position of the core and the smaller crosses those of components NW and SE. Beam sizes are approximately 4 arcsec$^{2}$. Artefacts caused by flare variability remain in images C10 through C13. It should be noted that C10 only includes data prior to the initial flare rise in order to reduce severity of artefacts.}
\end{minipage}
\end{figure} 

\clearpage

\begin{figure}
\begin{minipage}{7in}
\centerline{\includegraphics[width=6.8in]{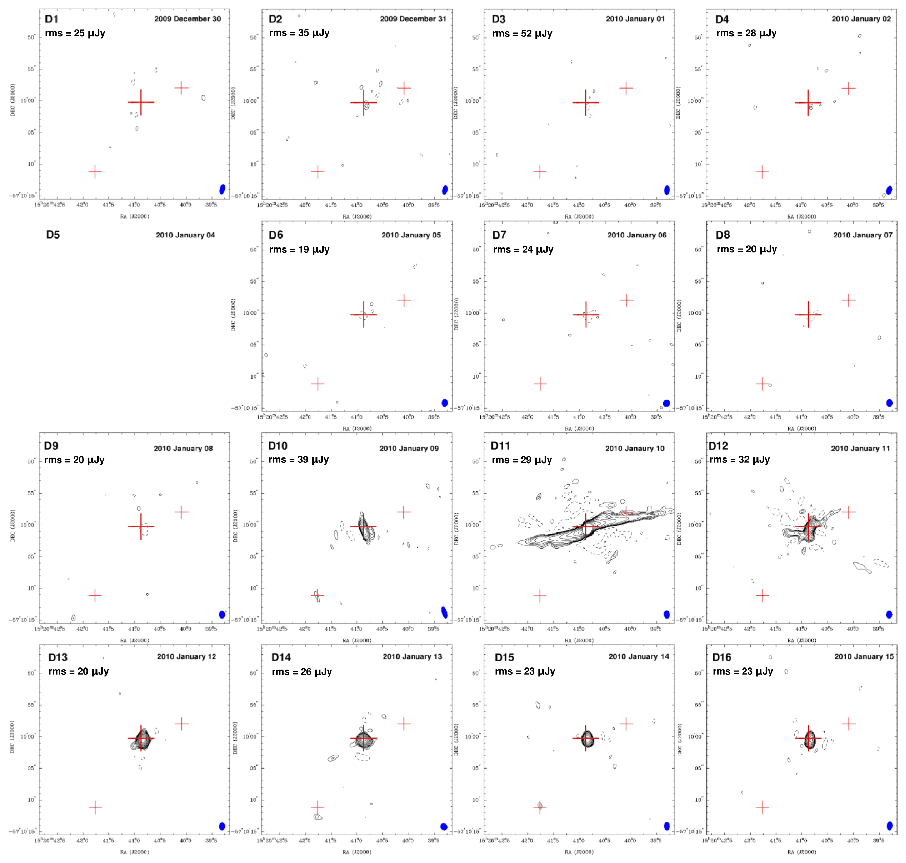}}
\caption{Difference radio maps of Circinus X-1 at 9 GHz. Each epoch has had data from 04 January 2010 subtracted prior to imaging (thus no image is available for 04 Jan itself). Weighting is determined with a robust factor of 0.5 (an optimal compromise between natural and uniform weighting). Contour lines are at -2.8, 2.8, 4, 5.6, 8, 11, 16, 23, 32, 45, 64, 90$\times$rms noise of each epoch (matching the scheme used in Tudose \etal 2008, rms listed in the top left corner of each panel). The larger cross marks the position of the core and the smaller crosses those of components NW and SE. Beam sizes are approximately 1.5 arcsec$^{2}$. Artefacts caused by flare variability remain in images D10 through D12. It should be noted that D10 only includes data prior to the initial flare rise in order to reduce severity of artefacts.}
\end{minipage}
\end{figure} 

\clearpage
\section{Model subtracted radio maps}

\begin{figure}
\centerline{\includegraphics[width=2.8in]{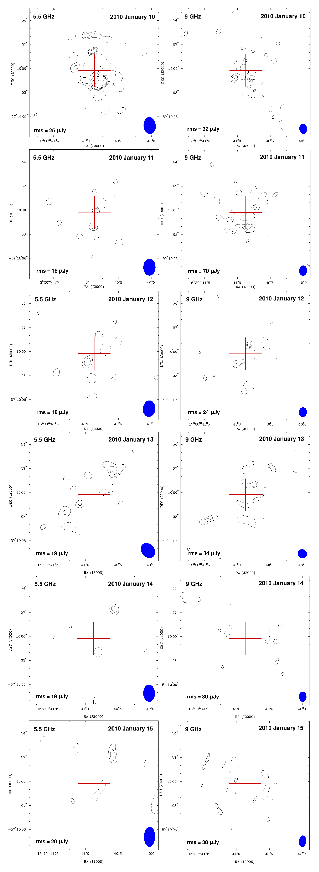}}
\caption{Model subtracted radio maps of Circinus X-1. These images have had both the Jan 04 data subtracted, and a model point source subtracted that is centred at the location of Cir X-1's core and whose behaviour is modelled on the core light curves. They cover the flare event and subsequent decay. Contour lines are at -2.8, 2.8, 4, 5.6, 8, 11, 16, 23, 32, 45, 64, 90$\times$rms noise of each epoch (listed on the bottom left of each panel), however the rms is difficult to measure accurately as image noise tends to increase towards Cir X-1's position, and thus values used are higher than those measured in the surrounding regions of maps. Beam sizes are approximately 4 arcsec$^{2}$.}
\end{figure}

\end{document}